\documentclass[twocolumn,pra,showpacs,superscriptaddress]{revtex4}
\usepackage{graphicx}

\begin{document}

\title{A nonlinear mechanism of charge qubit decoherence \\
in a lossy cavity: the quasi normal mode approach}
\author{Y. B. Gao}
\affiliation{Department of Applied Physics, Beijing University of Technology, Beijing,
100022, China}
\affiliation{Institute of Theoretical Physics, Chinese Academy of Sciences, Beijing,
100080, China}
\author{Y. D. Wang}
\affiliation{Institute of Theoretical Physics, Chinese Academy of Sciences, Beijing,
100080, China}
\author{C. P. Sun}
\affiliation{Institute of Theoretical Physics, Chinese Academy of Sciences, Beijing,
100080, China}
\affiliation{Department of Physics, Nankai University, Tianjin, 300071, China}
\date{\today}

\begin{abstract}
In the viewpoint of quasi normal modes, we describe a novel decoherence
mechanism of charge qubit of Josephson Junctions (JJ) in a lossy
micro-cavity, which can appear in the realistic experiment for quantum
computation based on JJ qubit. We show that the nonlinear coupling of a
charge qubit to quantum cavity field can result in an additional dissipation
of resonant mode due to its effective interaction between those non-resonant
modes and a resonant mode, which is induced by the charge qubit itself. We
calculate the characterized time of the novel decoherence by making use of
the system plus bath method.
\end{abstract}

\maketitle

\section{introduction}

The superposition principle is most basically governing quantum world \cite%
{Zurek83}. It is also the foundation of quantum information technology. The
ideal coherent superposition state can only be preserved in quantum world
isolated from external influences. However, the influence of the surrounding
environment can never be blocked off completely. Now the coherence is an
essential requirement for quantum information and the decoherence will
result in errors to reduce the power of quantum computation and quantum
communication or even destroys it. The quantum decoherence has become the
biggest obstacle to its implementation. In practice, we need a qubit with
long decoherence time and a longer lifetime medium to implement quantum
logic gate. To overcome quantum decoherence, one should know its dynamic
details theoretically and experimentally in various physical systems
including all kinds of qubits.

Recently in solid state based quantum computation, Josephson Junction (JJ)
qubits: charge qubit, flux qubit \cite{Makhlin01} or their hybridizations,
demonstrate their large potential as a candidate of scalable quantum
computation. On the one hand, the Rabi oscillation in a Cooper-pair box
(charge qubit) \cite{Nakamura99}, the existence of two-qubit states \cite%
{Pashkin99} and the entanglement between a flux qubit and a superconducting
quantum interference device (SQUID) \cite{Nakamura04} are realized
experimentally. Up to now, the decoherence time of JJ qubit has been the
order of 5$\mu s$ \cite{Han02}. On the other hand, to implement quantum
computation, one should integrate many qubits to form a quantum coherent
network. To this end, a longer lifetime medium is required to transfer
quantum information among these qubits in the network. Some investigations
have shown that quantized field in microwave cavity might be a good
candidate, whose lifetime is of order $1ms$ \cite{Raimond01}. For this
purpose, the integration of JJ qubit and cavity QED has become a focus in
exploring the JJ qubit based quantum computing \cite%
{Han03,You03,Saidi01,Liu04,Girvin03,Girvin04}.

In spite of these exciting advances, the relatively short coherent time is
still a problem in implementing the JJ based quantum computing in large
scale and the mechanism of decoherence of JJ qubit is not very clear,
especially in the presence of external field. The fluctuation of background
charge is a well-known source of quantum decoherence \cite{Nakamura02}, but
it is not the unique one. For a real dc-SQUID, the fluctuations of gate
voltage and magnetic flux produced by the screening current circulating
around the dc-SQUID may also cause additional quantum decoherence in the
charge qubit-cavity system. To bring out more clearly the physical mechanism
of nonlinear decoherence described in this paper, we can avoid the effect of
screening current in the physical case of the dc-SQUID screening parameter $%
\beta _{m}=2LI_{c}/\Phi _{0}\leq 2/\pi $ \cite{Tinkham96}. Here
$L$ is the loop inductance of dc-SQUID and $I_{c}$ the critical
current of Josephson junction and $\Phi _{0}$ flux quanta. To
simplify the analysis of our paper and demonstrate more clearly
the physics of our nonlinear decoherence mechanism, we do not
consider fluctuations of the gate voltage and other source of
decoherence. Most current investigations for integrating and
manipulating various kinds of JJ qubits mainly concern about the
idealized cavity without damping. Naturally one can question about
what will happen if we place a qubit in a non-ideal cavity. It is
our direct motivation for this paper. Here we will deal with
quantized field in a lossy cavity with the quasi normal mode
approach \cite{Fox1961,Lamb73,Scully97,Young98}. In this
treatment, cavity modes in a lossy cavity are divided into a
single resonant mode and other non-resonant multi-modes. Due to
the nonlinear coupling of the charge qubit to cavity field, the
effective interaction between those non-resonant modes and the
resonant one causes an additional dissipation of the resonant
mode. This is just the novel mechanism of quantum decoherence for
the charge qubits in a lossy cavity.

The paper is organized in the following sections. With the quasi normal mode
approach, the model in Sec. II describes a charge qubit interacting with a
lossy cavity. In Sec. III, we demonstrate how the nonlinear coupling of a
charge qubit to cavity field induces the effective interaction between
resonant mode and non-resonant ones. In Sec. IV, we find that the nonlinear
coupling leads to energy dissipation of resonant mode of cavity field which
is linked to quantum decoherence of a charge qubit in a lossy cavity in Sec.
V.

\section{Model: charge qubit coupled to the lossy cavity}

In this paper, we consider a single mode quantum field in a lossy cavity
with frequency $\omega \sim 30GHz$ (typically in the microwave domain) and
quality factor $Q\sim 10^{6}$ coupled to a charge qubit, in which the
charging energy $E_{C}\sim 122\mu eV$ and the Josephson coupling energy $%
E_{J}\sim 34\mu eV$ \cite{Nakamura02}. The charge qubit considered in this
paper is a dc-SQUID consisting of two identical Josephson junctions enclosed
by a superconducting loop. It is located in a lossy cavity which is produced
by a semi-transparent mirror. The similar case is discussed for quantum
dissipation of semiconductor exciton in a lossy cavity \cite{Liu01}. In this
paper, we can describe the magnetic field in a lossy cavity with the quasi
mode approach \cite{Fox1961,Lamb73,Scully97,Young98}. And we can divide
cavity modes into two parts: a single resonant mode and other non-resonant
ones.

\begin{figure}[th]
\includegraphics[width=5cm]{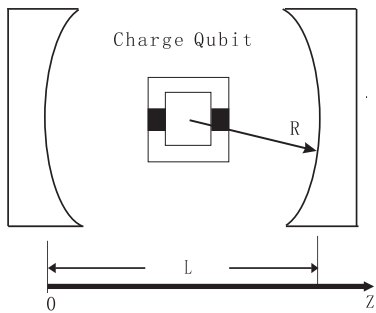}
\caption{Schematic of the charge qubit-cavity system. Superconducting
microwave cavity with parameter, $R=2.55$mm, $L=0.5$cm. }
\label{cavity}
\end{figure}

In this case, the magnetic flux threading the dc-SQUID is generated by
magnetic field $\mathbf{B}=\mathbf{B}_{c}+\mathbf{B}_{q}$, which consists of
external classical magnetic field $\mathbf{B}_{c}$ and quantum cavity field $%
\mathbf{B}_{q}$ \cite{You03}. To demonstrate the physics of our result, we
don't consider the effect of the screening current. Similarly the total
magnetic flux threading dc-SQUID could be a sum of two parts $\Phi =\Phi
_{c}+\Phi _{q}$, where $\Phi _{c}=\int \mathbf{B}_{c}\cdot d\mathbf{S}$ is
the external classical flux threading the dc-SQUID, $\Phi _{q}=\int \mathbf{B%
}_{q}\cdot d\mathbf{S}$ the cavity-induced quantum flux through the dc-SQUID
and S the area bounded by the dc-SQUID.

For an ideal cavity, we can describe cavity field with a set of normal modes
with different frequencies. Similarly as in a laser theory \cite{Lamb73}, we
adopt quasi mode approach to describe the quantum cavity field $\mathbf{B}%
_{q}$ in terms of a discrete set of quasi modes of the lossy cavity, each of
which (resonant mode) has a finite quality factor $Q$ and there exists many
modes of the universe (non-resonant modes) corresponding to each quasi mode.
Then the quantum cavity field $\mathbf{B}_{q}$ could be composed of resonant
mode of cavity field $\mathbf{B}_{r}$ and non-resonant modes of cavity field
$\mathbf{B}_{nr}$, $\mathbf{B}_{q}=\mathbf{B}_{r}+\mathbf{B}_{nr}$. In this
paper, we assume that the lossy cavity of our interest contains only one
quasi mode.

Fig.\ref{cavity} shows that the dc-SQUID lies in the x-z plane and the
direction of the quantum cavity field is perpendicular to the plane,
\[
B_{q,y}(z)=B_{r}(z)+B_{nr}(z)
\]%
and
\begin{eqnarray}
B_{r}(z) &=&-i\left( \frac{\hbar \omega }{\varepsilon _{0}Vc^{2}}\right) ^{%
\frac{1}{2}}\sin \left( \frac{\omega }{c}z\right) \left( a-a^{\dagger
}\right) ,  \nonumber \\
B_{nr}(z) &=&-i\sum_{j}M_{j}\left( \frac{\hbar \omega _{j}}{\varepsilon
_{0}V_{j}c^{2}}\right) ^{\frac{1}{2}}\times  \nonumber \\
&&\sin \left( \frac{\omega _{j}}{c}\left( z-L\right) \right) \left(
a_{j}-a_{j}^{\dag }\right) ,  \label{m-field-y}
\end{eqnarray}%
where $a^{\dagger }(a)$ and $a_{j}^{\dagger }(a_{j})$\ are the creation
(annihilation) operators corresponding to single resonant mode of frequency $%
\omega $ and non-resonant modes of frequencies $\omega _{j}$, $V$ ($V_{j}$)
the electromagnetic mode volume corresponding to resonant mode and
non-resonant mode respectively and $L$ the length of the cavity. The
constant $M_{j}$ in Eq.(\ref{m-field-y}) is proportional to a Lorentzian
\cite{Scully97}
\begin{equation}
M_{j}=\frac{\Lambda \frac{\gamma }{2}}{\sqrt{\left( \omega _{j}-\omega
\right) ^{2}+\left( \frac{\gamma }{2}\right) ^{2}}}  \label{lorentzian}
\end{equation}%
where $\gamma $ is the decay rate of a quasi mode of cavity , $\Lambda $ the
bandwidth associated with the cavity wall transparency and $\omega $ the
central frequency of the resonant mode of cavity. It is obvious that $M_{j}$
will acquire the maximum value when the frequency of non-resonant mode $%
\omega _{j}$ is very close to the central frequency of the resonant mode $%
\omega $. In our investigation, the lossy cavity contained only one quasi
mode.

The Hamiltonian for a charge qubit (dc-SQUID) can be written as in Ref. \cite%
{Makhlin99},
\begin{equation}
H=4E_{C}\left( n_{g}-\frac{1}{2}\right) \sigma _{z}-E_{J}\cos \left( \pi
\frac{\Phi }{\Phi _{0}}\right) \sigma _{x}.  \label{Ham1}
\end{equation}%
Where $E_{C}$ is the charging energy and $E_{J}$ the Josephson coupling
energy, $E_{C}>>E_{J}$ for a charge qubit. $\Phi $ is the magnetic flux
generated by controlled classical magnetic field and quantum cavity field
and $\Phi _{0}=h/2e$ the flux quanta. As a control parameter, the
dimensionless gate charge $n_{g}=C_{g}V_{g}/2e$ is determined by the gate
voltage $V_{g}$ applied on the gate capacitance $C_{g}$. Quasi-spin
operators
\[
\sigma _{z}=\left\vert 0\right\rangle _{q}\left\langle 0\right\vert
_{q}-\left\vert 1\right\rangle _{q}\left\langle 1\right\vert _{q},\sigma
_{x}=\left\vert 0\right\rangle _{q}\left\langle 1\right\vert _{q}+\left\vert
1\right\rangle _{q}\left\langle 0\right\vert _{q}
\]%
are defined in the charge qubit basis ($\left\vert 0\right\rangle _{q}$ and $%
\left\vert 1\right\rangle _{q}$).

In Fig.\ref{cavity}, the dc-SQUID is located at the position of the antinode
of standing wave field in cavity, i.e., $z=L/2$. Then corresponding to the
field decomposition $\mathbf{B}_{q}=\mathbf{B}_{r}+\mathbf{B}_{nr}$ the
magnetic flux $\Phi _{q}=\Phi _{r}+\Phi _{nr}$ enclosed by the dc-SQUID is
explicitly given by
\begin{eqnarray*}
\Phi _{r} &=&-iS\left( \frac{\hbar \omega }{\varepsilon _{0}Vc^{2}}\right) ^{%
\frac{1}{2}}\left( a-a^{\dagger }\right) , \\
\Phi _{nr} &=&-iS\sum_{j}M_{j}\left( \frac{\hbar \omega _{j}}{\varepsilon
_{0}V_{j}c^{2}}\right) ^{\frac{1}{2}}\sin \left( -\frac{\omega _{j}}{c}\frac{%
L}{2}\right) \left( a_{j}-a_{j}^{\dag }\right) .
\end{eqnarray*}%
In a straightforward way, we derive the Hamiltonian of the qubit-cavity
system from Eq.(\ref{Ham1}),
\begin{eqnarray}
H &=&4E_{C}\left( n_{g}-\frac{1}{2}\right) \sigma _{z}-E_{J}\cos \left( \phi
_{c}+\phi _{q}\right) \sigma _{x}  \nonumber \\
&&+\hbar \omega a^{\dag }a+\sum\limits_{j}\hbar \omega _{j}a_{j}^{\dag }a_{j}%
\text{.}  \label{Ham}
\end{eqnarray}%
Where $\phi _{c}$ and $\phi _{q}$ are phase and "phase operator" generated
by the flux $\Phi _{c}$ and $\Phi _{q}$ respectively,
\begin{eqnarray*}
\phi _{c} &=&\frac{\pi \Phi _{c}}{\Phi _{0}}, \\
\phi _{q} &=&-i\phi _{0}\left( a-a^{\dag }\right) -i\sum_{j}\phi _{j}\left(
a_{j}-a_{j}^{\dag }\right)
\end{eqnarray*}%
and
\begin{eqnarray*}
\phi _{0} &=&\frac{\pi S}{\Phi _{0}}\left( \frac{\hbar \omega }{\varepsilon
_{0}Vc^{2}}\right) ^{\frac{1}{2}}, \\
\phi _{j} &=&M_{j}\frac{\pi S}{\Phi _{0}}\left( \frac{\hbar \omega _{j}}{%
\varepsilon _{0}V_{j}c^{2}}\right) ^{\frac{1}{2}}\sin \left( -\frac{\omega
_{j}}{c}\frac{L}{2}\right) .
\end{eqnarray*}

In the above discussions to achieve the simplified model we have ignored the
effect of the screening current. However, if the inductance of the loop of
dc-SQUID is not zero, the screening current will induce the additional
decoherence. In presence of the screening current, we can not neglect the
difference between the practical magnetic flux $\Phi $ threading the
dc-SQUID and the external magnetic flux $\Phi _{x}$. It can be determined by
the following equation \cite{Tinkham96}%
\[
\phi =\phi _{x}+\frac{\pi }{2}\beta _{m}\sin \phi
\]%
where $\beta _{m}=2LI_{c}/\Phi _{0}$ is usually called screening parameter
and $\phi =\pi \Phi /\Phi _{0}$ and $\phi _{x}=\pi \Phi _{x}/\Phi _{0}$.
This equation simply shows that the total flux $\Phi $ is the sum of the
external flux and the induced flux determined only by $\Phi $ itself. If the
screening parameter $\beta _{m}$ is small enough, we can approximately solve
this equation in a single value domain with the technique of perturbation
recursion up to the second order%
\begin{equation}
\phi =\phi _{x}+\frac{\pi }{2}\beta _{m}\sin \phi _{x}+\left( \frac{\pi }{2}%
\beta _{m}\right) ^{2}\sin \phi _{x}\cos \phi _{x}\text{.}  \label{s-phi}
\end{equation}%
When $\beta _{m}\pi /2<<1$, i.e., $\beta _{m}<<2/\pi $, we can ignore the
effect of the screening current. But the additional nonlinear terms
containing $\left( \phi _{x}\right) ^{2}$ at least should also induce the
additional nonlinear interaction between the resonant mode and the
non-resonant ones. This is the further result in the decoherence of charge
qubit in the lossy cavity. However, to clearly demonstrate the physics of
the central results of our paper, we suppose $\beta _{m}<<2/\pi $ and ignore
the effect of the screening current. Therefore we do not give much details
for this problem.

\section{Mode Interaction induced by nonlinear coupling to cavity}

As shown in Fig.\ref{cavity}, two spherical mirrors form microwave cavity
\cite{Raimond01} containing a single mode standing wave field. And an
external classical magnetic field is also injected into the cavity. In this
paper, the geometry of cavity is described by the parameters: the curvature
radius $R=2\,.55$mm, the width between two mirrors $L=0.5$cm. By some
straightforward calculations, we get that the cavity field $B=\left( \hbar
\omega /\epsilon _{0}Vc^{2}\right) ^{\frac{1}{2}}=7.52\times 10^{-11}$%
(Tesla) and $\phi _{0}=\pi \Phi _{q}/\Phi _{0}=1.14\times 10^{-5}$. In a low
photon number cavity, we find that $\phi _{q}\ll \phi _{c}$, thus there is
only a weak polynomial nonlinearity in Eq.(\ref{Ham}).

To simplify the Hamiltonian in Eq.(\ref{Ham}), we expand $\cos \left( \phi
_{c}+\phi _{q}\right) $ in terms of small quantity $\phi _{q}$ up to the
second order,
\begin{equation}
\cos \left( \phi _{c}+\phi _{q}\right) =\left( 1-\frac{1}{2}\phi
_{q}^{2}\right) \cos \phi _{c}-\phi _{q}\sin \phi _{c}
\label{non-linear coupling}
\end{equation}
Obviously we can know that the second order term $\phi _{q}^{2} =(\phi
_{r}+\phi _{nr})^2$ includes the term $\phi _{r}\phi _{nr}$ which results in
the nonlinear coupling between resonant mode and non-resonant modes of
cavity field, on which the results of this paper is based. The first order
term $\phi _{q}$ is linearly dependent of $\phi _{r}$ and $\phi _{nr}$,
which can not lead to the coupling between resonant mode and non-resonant
ones. Therefore the nonlinear coupling in terms of $\phi _{q}^{2}\sim \phi
_{r}\phi _{nr}$ will induce energy dissipation and quantum decoherence of
the charge qubit in a lossy cavity simultaneously.

To clearly demonstrate the effect of quantum decoherence of a charge qubit
in a lossy cavity we tune the gate voltage $V_{g}$ such that $n_{g}=1/2$ to
eliminate the effect of background charge fluctuation up to the linear
order. Then the effective Hamiltonian corresponding to a standard quantum
measurement model \cite{Wang02} reads
\begin{equation}
H=H^{\left( 0\right) }\left\vert 0\right\rangle \left\langle 0\right\vert
+H^{\left( 1\right) }\left\vert 1\right\rangle \left\langle 1\right\vert ,
\label{measure-Ham}
\end{equation}%
which is diagonal with respect to eigenstates of quasi spin operator $%
\sigma_{x}$, $\left\vert 0\right\rangle =\left\vert 0\right\rangle
_{q}+\left\vert 1\right\rangle _{q}$ and $\left\vert 1\right\rangle
=\left\vert 0\right\rangle _{q}-\left\vert 1\right\rangle _{q} $.

As seen in Eq.(\ref{non-linear coupling}), the second order term $\phi
_{q}^{2}\sim \phi _{r}\phi _{nr}$ results in the interaction $\sim \left(
a_{j}-a_{j}^{\dag }\right) \left( a-a^{\dag }\right) $ between single
resonant mode and other non-resonant ones, while the first order term $\phi
_{q}$ results in the forced terms $\left( a-a^{\dag }\right) $ and $\left(
a_{j}-a_{j}^{\dag }\right) $ in the above Hamiltonian. With the rotating
wave approximation (RWA), we can drop down the terms of $a^{2}$ ($a^{\dag 2}$%
) and $a_{j}a$ ($a_{j}^{\dag }a^{\dag }$) in $\cos \left( \phi _{c}+\phi
_{q}\right) $ and get an effective Hamiltonian
\begin{equation}
H^{\left( k\right) }=H_{s}^{\left( k\right) }+H_{I}^{\left( k\right)
}+H_{B}^{\left( k\right) }+N^{(k)}  \label{forced-Ham}
\end{equation}%
where
\begin{eqnarray*}
H_{s}^{\left( k\right) } &=&\hbar \Omega ^{\left( k\right) }a^{\dag }a-i\xi
^{\left( k\right) }\left( a-a^{\dag }\right) , \\
H_{I}^{\left( k\right) } &=&\sum\limits_{j}g_{j}^{\left( k\right) }\left(
a_{j}a^{\dag }+a_{j}^{\dag }a\right) , \\
H_{B}^{\left( k\right) } &=&\sum\limits_{j}\hbar \omega _{j}a_{j}^{\dag
}a_{j}-i\sum_{j}\xi _{j}^{\left( k\right) }\left( a_{j}-a_{j}^{\dag }\right)
\end{eqnarray*}%
and the parameters in the above equation can be explicitly expressed as
\begin{eqnarray}
N^{(k)} &=&\frac{\left( -1\right) ^{k}}{\hbar }\left( \phi _{0}^{2}E_{J}\cos
\phi _{c}-E_{J}\cos \phi _{c}\right) ,  \nonumber \\
\Omega ^{\left( k\right) } &=&\omega +\frac{\left( -1\right) ^{k}}{\hbar }%
\phi _{0}^{2}E_{J}\cos \phi _{c},  \nonumber \\
g_{j}^{\left( k\right) } &=&\frac{\left( -1\right) ^{k}}{\hbar }\phi
_{j}\phi _{0}E_{J}\cos \phi _{c},  \nonumber \\
\xi ^{\left( k\right) } &=&\frac{\left( -1\right) ^{k}}{\hbar }\phi
_{0}E_{J}\sin \phi _{c},  \nonumber \\
\xi _{j}^{\left( k\right) } &=&\frac{\left( -1\right) ^{k}}{\hbar }\phi
_{j}E_{J}\sin \phi _{c}  \label{parameter-1}
\end{eqnarray}%
for $k=0,1$. Here $H_{s}^{\left( k\right) }$ describes a system with a
forced oscillator of frequency $\Omega ^{\left( k\right) }$, $H_{B}^{\left(
k\right) }$ describes a bath of many forced oscillators of frequency $\omega
_{j}$s, and $H_{I}^{\left( k\right) }$ describes the coupling of the
resonant mode to non-resonant modes. The coupling constant $g_{j}^{\left(
k\right) }$ owns a Lorentz type factor, i.e.,
\[
g_{j}^{\left( k\right) }\sim \frac{1}{\sqrt{\left( \omega _{j}-\omega
\right) ^{2}+\left( \frac{\gamma }{2}\right) ^{2}}}.
\]%
It means that the resonant mode of cavity field dominates the strength of
the interaction mostly.

\section{quantum dissipation of resonant mode of cavity field}

In this section, we study quantum dissipation of the resonant mode of cavity
field. In each component of the Hamiltonian in Eq.(\ref{forced-Ham}), $%
H^{\left( k\right) }$ can result in quantum dissipation of the resonant
mode. To solve dynamic equation governed by the effective Hamiltonian $%
H^{\left( k\right) }$, we rewrite the above Hamiltonian into the new form%
\begin{equation}
H^{\left( k\right) }=\hbar \Omega ^{\left( k\right) }b^{\dag
}b+\sum\limits_{j}\hbar \omega _{j}b_{j}^{\dag
}b_{j}+\sum\limits_{j}g_{j}^{\left( k\right) }\left( b_{j}b^{\dag
}+b_{j}^{\dag }b\right) +\varphi _{k}  \label{b-Ham}
\end{equation}%
by defining a new set of bosonic operators $b$ ($b^{\dag }$) and $b_{j}$ ($%
b_{j}^{\dag }$), which are the displacements of operators $a$ and $a_{j}$,
\begin{eqnarray}
b &=&a+\lambda ,  \label{new-form} \\
b_{j} &=&a_{j}+\lambda _{j}.  \nonumber
\end{eqnarray}%
Here, $\varphi _{k}$ is the constant, $\lambda $ and $\lambda _{j}$ are
dependent of the forced terms $\left( a-a^{\dag }\right) $ and $\left(
a-a^{\dag }\right) $ of the effective Hamiltonian $H^{\left( k\right) }$ in
Eq.(\ref{forced-Ham}).

For any coherent state $\left\vert \alpha \right\rangle _{a}$ and $%
\left\vert \alpha _{j}\right\rangle _{a_{j}}$ defined with respect to
annihilation operators $a$ and $a_{j}$, we get coherent state $\left\vert
\alpha \right\rangle _{b}$ and $\left\vert \alpha _{j}\right\rangle _{b_{j}}$
defined with respect to annihilation operators $b$ and $b_{j}$,
\begin{eqnarray}
\left\vert \alpha \right\rangle _{b} &=&e^{-\lambda \alpha ^{\ast
}}\left\vert \alpha -\lambda \right\rangle _{a},  \nonumber \\
\left\vert \alpha _{j}\right\rangle _{b_{j}} &=&e^{-\lambda _{j}\alpha
_{j}^{\ast }}\left\vert \alpha _{j}-\lambda _{j}\right\rangle _{a_{j}}.
\label{co-state-1}
\end{eqnarray}

Obviously we can see that the effective Hamiltonian $H^{\left( k\right) }$
in Eq.(\ref{b-Ham}) describes a typical dissipative system of a single-mode
boson soaked in a bath of many bosons (we have studied its wave function
structure in details \cite{Sun98}). The well-known solutions of Heisenberg
equations for the Hamiltonian $H^{\left( k\right) }$ is given in Ref. \cite%
{Sun98},
\begin{eqnarray*}
b^{\left( k\right) }\left( t\right) &=&u^{\left( k\right) }\left( t\right)
b\left( 0\right) +\sum_{j}v_{j}^{\left( k\right) }\left( t\right)
b_{j}\left( 0\right) , \\
b_{j}^{\left( k\right) }\left( t\right) &=&e^{-i\omega _{j}t}b_{j}\left(
0\right) +u_{j}^{\left( k\right) }\left( t\right) b\left( 0\right)
+\sum_{s}v_{j,s}^{\left( k\right) }\left( t\right) b_{s}\left( 0\right) .
\end{eqnarray*}%
Where $b^{\left( k\right) }\left( t\right) $ and $b_{j}^{\left( k\right)
}\left( t\right) $ represent the time evolution of operators $b$ and $b_{j}$%
. And we also get the solutions of Heisenberg equations for $a^{\left(
k\right) }\left( t\right) $ and $a_{j}^{\left( k\right) }\left( t\right) $
representing the time evolution of operator $a$ and $a_{j}$ driven by the
Hamiltonian $H^{\left( k\right) }$ as
\begin{eqnarray}
a^{\left( k\right) }\left( t\right) &=&b^{\left( k\right) }\left( t\right)
-\lambda ,  \nonumber \\
a_{j}^{\left( k\right) }\left( t\right) &=&b_{j}^{\left( k\right) }\left(
t\right) -\lambda _{j},  \label{displacement}
\end{eqnarray}%
where
\begin{eqnarray}
u^{\left( k\right) }\left( t\right) &=&e^{-\frac{\gamma }{2}t}e^{-i\left(
\Omega ^{\left( k\right) }+\Delta \Omega ^{\left( k\right) }\right) t},
\nonumber \\
v_{j}^{\left( k\right) }\left( t\right) &=&-\frac{g_{j}^{\left( k\right)
}e^{-i\omega _{j}t}\left( 1-e^{-i\left( \Omega ^{\left( k\right) }+\Delta
\Omega ^{\left( k\right) }-\omega _{j}\right) t}e^{-\frac{\gamma }{2}%
t}\right) }{\Omega ^{\left( k\right) }+\Delta \Omega ^{\left( k\right)
}-\omega _{j}-i\frac{\gamma }{2}},  \nonumber \\
u_{j}^{\left( k\right) }\left( t\right) &=&-\frac{g_{j}^{\left( k\right)
}e^{-i\omega _{j}t}\left( 1-e^{-i\left( \Omega ^{\left( k\right) }+\Delta
\Omega ^{\left( k\right) }-\omega _{j}\right) t}e^{-\frac{\gamma }{2}%
t}\right) }{\Omega ^{\left( k\right) }+\Delta \Omega ^{\left( k\right)
}-\omega _{j}-i\frac{\gamma }{2}},  \nonumber \\
v_{j,s}^{\left( k\right) }\left( t\right) &=&-\frac{g_{j}^{\left( k\right)
}g_{s}^{\left( k\right) }e^{-i\omega _{j}t}}{\Omega ^{\left( k\right)
}+\Delta \Omega ^{\left( k\right) }-\omega _{s}-i\frac{\gamma }{2}}\times
\nonumber \\
&&\left( \frac{1-e^{-i\left( \Omega ^{\left( k\right) }+\Delta \Omega
^{\left( k\right) }-\omega _{j}\right) t}e^{-\frac{\gamma }{2}t}}{\Omega
^{\left( k\right) }+\Delta \Omega ^{\left( k\right) }-\omega _{j}-i\frac{%
\gamma }{2}}+\Lambda \right) ,  \nonumber \\
\Lambda &=&t\text{, when }s=j\text{,}  \nonumber \\
\Lambda &=&\frac{e^{-i\left( \omega _{s}-\omega _{j}\right) t}-1}{\omega
_{s}-\omega _{j}}\text{, when }s\neq j\text{.}  \label{parameter-0}
\end{eqnarray}%
Where $\Delta \Omega ^{\left( k\right) }$ is frequency shift of $\Omega
^{\left( k\right) }$ corresponding to two different Hamiltonians $H^{\left(
k\right) }$. In general, $\Delta \Omega ^{\left( k\right) }$ can be absorbed
into $\Omega ^{\left( k\right) }$, i.e., $\Omega ^{\left( k\right) }\sim
\Omega ^{\left( k\right) }+\Delta \Omega ^{\left( k\right) }$. If the forced
terms of the Hamiltonian in Eq.(\ref{forced-Ham}) are absent, we will get
that $a^{\left( k\right) }\left( t\right) =b^{\left( k\right) }\left(
t\right) $ and $a_{j}^{\left( k\right) }\left( t\right) =b_{j}^{\left(
k\right) }\left( t\right) $.

Here we calculate the time evolution of the mean photon number of the
resonant mode of cavity field corresponding to two different Hamiltonians $%
H^{\left( k\right) }$. When we assume that the initial state of all modes of
cavity field is prepared in Fock state $\left\vert n,\left\{ n_{j}\right\}
\right\rangle _{a}=\left\vert n\right\rangle _{a}\otimes \left\vert \left\{
n_{j}\right\} \right\rangle _{a}$, the mean photon number of resonant mode
corresponding to $H_{k}$ is calculated as
\begin{eqnarray}
n^{\left( k\right) }\left( t\right) &=&\left\langle a^{\left( k\right) \dag
}\left( t\right) a^{\left( k\right) }\left( t\right) \right\rangle  \nonumber
\\
&=&e^{-\gamma t}n+F\left( \lambda ,\left\{ \lambda _{j}\right\} ,\left\{
n_{j}\right\} ,t\right)  \label{average-number}
\end{eqnarray}%
where the time dependent constant $F\left( \lambda ,\lambda
_{j},n_{j},t\right) $ is dependent of $\lambda $, $\lambda _{j}$ and $n_{j}$
and the time dependent term $n\exp \left( -\gamma t\right) $ characterizing
the quantum dissipation induced by non-resonant modes. Thus we can know that
the time evolution of mean number of the resonant mode $n^{\left( k\right)
}\left( t\right) $ is the sum of two parts: (1) quantum dissipation $n\exp
\left( -\gamma t\right) $ induced by non-resonant modes; (2) the constant $%
F\left( \lambda ,\left\{ \lambda _{j}\right\} ,\left\{ n_{j}\right\}
,t\right) $ generated by the first order term $\phi _{q}$ and mean number of
non-resonant modes $n_{j}$.

Through some simple calculations, we find that the constant $F\left( \lambda
,\left\{ \lambda _{j}\right\} ,\left\{ n_{j}\right\} ,t\right) $ will
approach zero when the forced terms vanish and the initial state of the bath
(non-resonant modes) in vacuum state $\left\vert \left\{ n_{j}\right\}
\right\rangle =\left\vert \left\{ 0_{j}\right\} \right\rangle $. Then we get
the same results of Ref. \cite{Louisell73} that the time evolution of mean
photon number of resonant mode is
\begin{equation}
n^{\left( k\right) }\left( t\right) =ne^{-\gamma t}.
\label{dissipation-mode}
\end{equation}%
It means that vacuum fluctuation of non-resonant modes leads to quantum
dissipation of resonant mode when the forced terms of operators $a$ and $%
a_{j}$ are absent. In the other words, the nonlinear coupling directly
causes quantum dissipation of the resonant mode in a lossy cavity.

In contrast to the model of single boson interacting with a bath of many
bosons, the constant $F\left( \lambda ,\lambda _{j},n_{j},t\right) $
provides the different effect that the mean number of resonant mode of
cavity field doesn't approach zero when time $t\rightarrow \infty $.

\section{decoherence induced by the dissipation of resonant mode}

Technically the process of quantum decoherence is described by the time
evolution of the reduced density matrix of the coupled qubit-cavity system.
To analyze it, we can calculate reduced density matrix for the time
evolution of the charge qubit. The pure decoherence process means that the
off diagonal elements of the reduced density matrix of the qubit vanish,
while the diagonal elements remain unchanged in an ideal case.

Now if the initial state of cavity field is in coherent state, i.e., the
resonant mode of cavity field is in a coherent state $\left\vert \alpha
\right\rangle $ and other non-resonant modes of cavity field in coherent
state $\left\vert \left\{ \alpha _{j}\right\} \right\rangle $, the initial
state of the total qubit-cavity system can be written as
\[
\left\vert \Psi \left( 0\right) \right\rangle =\left( C_{0}\left\vert
0\right\rangle +C_{1}\left\vert 1\right\rangle \right) \otimes \left\vert
\alpha ,\left\{ \alpha _{j}\right\} \right\rangle _{a}.
\]%
Then we can easily get the time evolution of the wave function for the
qubit-cavity system%
\begin{eqnarray}
\left\vert \Psi \left( t\right) \right\rangle &=&U\left( t\right) \left\vert
\Psi \left( 0\right) \right\rangle  \nonumber \\
&=&C_{0}\left\vert 0\right\rangle \otimes \left\vert \varphi ^{\left(
0\right) }\left( t\right) \right\rangle +C_{1}\left\vert 1\right\rangle
\otimes \left\vert \varphi ^{\left( 1\right) }\left( t\right) \right\rangle
\label{evolution}
\end{eqnarray}%
where $U^{\left( k\right) }\left( t\right) =\exp \left( -iH^{\left( k\right)
}t\right) $ are evolution operators for the effective Hamiltonians $%
H^{\left( k\right) }$ and
\begin{equation}
\left\vert \varphi ^{\left( k\right) }\left( t\right) \right\rangle
=U^{\left( k\right) }\left( t\right) \left\vert \alpha ,\left\{ \alpha
_{j}\right\} \right\rangle _{a}.  \label{k-state}
\end{equation}%
Then we can obtain time evolution of density matrix for the qubit-cavity
system
\begin{equation}
\rho \left( t\right) =\left\vert \psi \left( t\right) \right\rangle
\left\langle \psi \left( t\right) \right\vert  \label{density}
\end{equation}%
and calculate the reduced density matrix of the qubit
\begin{eqnarray}
\rho _{s}\left( t\right) &=&C_{0}^{\ast }C_{0}\left\vert 0\right\rangle
\left\langle 0\right\vert +C_{1}^{\ast }C_{1}\left\vert 1\right\rangle
\left\langle 1\right\vert  \nonumber \\
&&+\left\langle \varphi ^{\left( 1\right) }\left( t\right) |\varphi ^{\left(
0\right) }\left( t\right) \right\rangle C_{1}^{\ast }C_{0}\left\vert
0\right\rangle \left\langle 1\right\vert +h.c.  \label{reduced density}
\end{eqnarray}%
As a measure of the coherence of quantum system \cite{Sun93}, the
decoherence factor of charge qubit can be calculated as%
\begin{equation}
D\left( t\right) =\left\vert \left\langle \varphi ^{\left( 1\right) }\left(
t\right) |\varphi ^{\left( 0\right) }\left( t\right) \right\rangle
\right\vert .  \label{d-factor}
\end{equation}

For any coherent state, we have%
\[
\left\vert \alpha \right\rangle _{b}=\exp \left[ \alpha b^{\dag }\left(
0\right) -\alpha ^{\ast }b\left( 0\right) \right] \left\vert 0\right\rangle
_{b}.
\]%
Obviously we can see that
\[
U^{\left( k\right) }\left( t\right) \left\vert 0,\left\{ 0_{j}\right\}
\right\rangle _{b}=0.
\]

Formally, we can define
\[
O^{\left( k\right) }\left( t\right) =U^{\left( k\right) }\left( t\right)
O\left( 0\right) U^{\left( k\right) \dag }\left( t\right)
\]%
for any operator $O^{\left( k\right) }$ corresponding to $H^{\left( k\right)
}$ and get the time dependent equation%
\begin{equation}
\frac{dO^{\left( k\right) }\left( t\right) }{dt}=i\left[ O^{\left( k\right)
}\left( t\right) ,H_{k}\right] .  \label{anti-Heisenberg}
\end{equation}%
So we have $A^{\left( k\right) }\left( t\right) =U^{\left( k\right) }\left(
t\right) b\left( 0\right) U^{\left( k\right) \dag }\left( t\right) $ and $%
A_{j}^{\left( k\right) }\left( t\right) =U^{\left( k\right) }\left( t\right)
b_{j}\left( 0\right) U^{\left( k\right) \dag }\left( t\right) $. By
substituting $-i$ into Heisenberg equation with $i$, we can get the
solutions of Eq.(\ref{anti-Heisenberg})
\begin{eqnarray*}
A^{\left( k\right) }\left( t\right) &=&u^{\left( k\right) \ast }\left(
t\right) b\left( 0\right) +\sum_{j}v_{j}^{\left( k\right) \ast }\left(
t\right) b_{j}\left( 0\right) , \\
A_{j}^{\left( k\right) }\left( t\right) &=&e^{i\omega _{j}t}b_{j}\left(
0\right) +u_{j}^{\left( k\right) \ast }\left( t\right) b\left( 0\right)
+\sum_{s}v_{j,s}^{\left( k\right) \ast }\left( t\right) b_{s}\left( 0\right)
.
\end{eqnarray*}

To demonstrate the effect of quantum dissipation of resonant mode induced by
non-resonant modes on decoherence of charge qubit, here we do not consider
the effect of the forced terms and set $\lambda =0$ and $\lambda _{j}=0$.

In the above section, we have known that the vacuum fluctuation induced by
the non-resonant modes can result in quantum dissipation of resonant mode.
So we assume that the initial state of non-resonant modes is in vacuum state
$\left\vert \left\{ 0_{j}\right\} \right\rangle _{b}=\left\vert \left\{
0_{j}\right\} \right\rangle _{a}$ and initial state of resonant mode
coherent state $\left\vert \alpha \right\rangle _{b}=\left\vert \alpha
\right\rangle _{a}$. And time evolution of decoherence factor of the charge
qubit is
\begin{equation}
D\left( t\right) =e^{-\frac{\left\vert \alpha \right\vert ^{2}}{2}\left(
\left\vert u^{\left( 1\right) }\left( t\right) -u^{\left( 0\right) }\left(
t\right) \right\vert ^{2}+\sum_{j}\left\vert v_{j}^{\left( 1\right) }\left(
t\right) -v_{j}^{\left( 0\right) }\left( t\right) \right\vert ^{2}\right) }.
\label{simplify-d-factor}
\end{equation}

\begin{figure}[th]
\includegraphics[width=6.5cm]{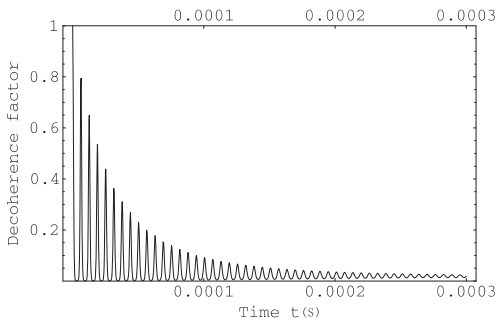}
\caption{Decoherence factor $D\left( t\right) $ in Eq.(\protect\ref%
{deco-final}) as the function of the time $t$ with the value of $\protect%
\alpha =2$.}
\label{deco}
\end{figure}

With commutation relation $\left[ A^{\left( k\right) }\left( t\right)
,A^{\left( k\right) \dag }\left( t\right) \right] =1$, we have
\begin{equation}
\left\vert u^{\left( k\right) }\left( t\right) \right\vert
^{2}+\sum_{j}\left\vert v_{j}^{\left( k\right) }\left( t\right) \right\vert
^{2}=1.  \nonumber
\end{equation}%
According to Eq.(\ref{k-state}), we have%
\begin{equation}
\left\vert \varphi _{k}\left( t\right) \right\rangle =\left\vert \alpha
u^{\left( k\right) }\left( t\right) \right\rangle _{b}\otimes \Pi
_{j}\left\vert \alpha v_{j}^{\left( k\right) }\left( t\right) \right\rangle
_{b_{j}}.  \nonumber
\end{equation}%
And the decoherence factor in Eq.(\ref{simplify-d-factor}) becomes
\begin{eqnarray}
D\left( t\right) &=&e^{-\left\vert \alpha \right\vert ^{2}+\frac{\left\vert
\alpha \right\vert ^{2}}{2}\left( u^{\left( 1\right) \ast }\left( t\right)
u^{\left( 0\right) }\left( t\right) +u^{\left( 1\right) }\left( t\right)
u^{\left( 0\right) \ast }\left( t\right) \right) }  \nonumber \\
&&\times e^{\frac{\left\vert \alpha \right\vert ^{2}}{2}\sum_{j}\left(
v_{j}^{\left( 1\right) \ast }\left( t\right) v_{j}^{\left( 0\right) }\left(
t\right) +v_{j}^{\left( 1\right) }\left( t\right) v_{j}^{\left( 0\right)
\ast }\left( t\right) \right) }.  \label{simplify-d-factor-1}
\end{eqnarray}

From the above results in Eq.(\ref{parameter-1}) and Eq.(\ref{parameter-0}),
we know that the term $v_{j}^{\left( 1\right) \ast }\left( t\right)
v_{j}^{\left( 0\right) }\left( t\right) $ is proportional to $\phi
_{j}^{2}\phi _{0}^{2}\sim \phi _{0}^{4}$ which is a 4th order term of
nonlinear expansion in Eq.(\ref{non-linear coupling}). Then we can omit it
in the calculation of the decoherence factor and get
\begin{equation}
D\left( t\right) =e^{-\left\vert \alpha \right\vert ^{2}\left( 1-e^{-\gamma
t}\cos \left[ \left( \Omega ^{\left( 1\right) }+\Delta \Omega ^{\left(
1\right) }-\Omega ^{\left( 0\right) }-\Delta \Omega ^{\left( 0\right)
}\right) t\right] \right) }.  \label{deco-final}
\end{equation}%
Fig.\ref{deco} shows that the decoherence factor $D\left( t\right) $ will
decrease in the oscillating decay form. In the above equation, the term $%
\exp \left( -\gamma t\right) $ represents quantum dissipation induced by the
non-resonant modes. Therefore we obtain the central result of this paper
that quantum dissipation of the resonant mode induced by the non-resonant
modes directly results in quantum decoherence of the charge qubit in the
lossy cavity. At long times, the decoherence factor $D\left( t\right) =\exp
\left( -\left\vert \alpha \right\vert ^{2}\right) $ is determined by the
mean photon number $\left\vert \alpha \right\vert ^{2}$. At short times, the
decoherence factor becomes $D\left( t\right) =\exp \left( -\Gamma t\right) $%
. Where the decay rate of decoherence $\Gamma $ is proportional to the mean
photon number of resonant mode $\left\vert \alpha \right\vert ^{2}$ and
decay rate $\gamma $ of quantum dissipation of resonant mode, $\Gamma
=\gamma \left\vert \alpha \right\vert ^{2}$.

\section{conclusions}

Before concluding this paper, we would like to remark the influences of the
fluctuations of gate charge $n_{g}$ around $1/2$. We notice that the
classical fluctuation of gate voltage is not the unique source of
decoherence. The most recent investigations have demonstrated that $1/f$
noise due to the background charge fluctuation, which also plays an
important role in the decoherence of charge qubit \cite%
{Nakamura02,Faoro02,Faoro04}. Maybe there also exists some unknown source of
decoherence. For simplicity we only deal with the decoherence of charge
qubit by considering that such fluctuations can indeed occur in a real
dc-SQUID, but it may be ignored so as to bring out more clearly the novel
mechanism. Our investigation only emphasizes the role that nonlinear
coupling plays in the decoherence of charge qubit in some case.

In this paper, we have discovered the novel phenomenon of quantum
decoherence of charge qubit in a lossy cavity, where we adopt the quasi mode
approach to deal with a lossy cavity. We find that the nonlinear coupling
between the charge qubit and the cavity field can induce the interaction
between resonant mode and non-resonant modes of cavity field. Based on this
observation, we achieve a nice model for this novel decoherence mechanism
that a forced oscillator (resonant mode) interacts with a bath of many
forced oscillators (non-resonant modes). The decoherence factor is
calculated to demonstrate an oscillating decay of quantum coherence of a
charge qubit in the lossy cavity. In addition, we have shown that vacuum
fluctuation provided by these non-resonant modes can cause the quantum
dissipation of resonant mode. Consequently, the quantum dissipation of
resonant mode directly results in quantum decoherence of charge qubit. These
analysis describe the novel source of quantum decoherence for a charge qubit
in the lossy cavity.

\begin{acknowledgments}
We thank the support of the CNSF (grant No.90203018) and the Knowledge
Innovation Program (KIP) of the Chinese Academy of Sciences and the National
Fundamental Research Program of China with No.001GB309310. We also sincerely
thank C. Li, Y. Li and P. Zhang for helpful discussions.
\end{acknowledgments}

\end{document}